\documentclass[aps,preprint]{revtex4}%
\usepackage{amsfonts}
\usepackage{amsmath}
\usepackage{amssymb}
\usepackage{graphicx}%
\providecommand{\U}[1]{\protect\rule{.1in}{.1in}}
\begin{document}
\title{Conditional probabilities and collapse in quantum measurements}
\author{Roberto Laura}
\affiliation{Departamento de F\'{\i}sica y Qu\'{\i}mica, F.C.E.I.A., Universidad Nacional
de Rosario. Av. Pellegrini 250, 2000 Rosario, Argentina. e-mail: rlaura@fceia.unr.edu.ar}
\author{Leonardo Vanni}
\affiliation{Instituto de Astronom\'{\i}a y F\'{\i}sica del Espacio. Casilla de Correos 67,
Sucursal 28, 1428 Buenos Aires, Argentina. e-mail: lv@iafe.uba.ar}

\begin{abstract}
We show that including both the system and the apparatus in the quantum
description of the measurement process, and using the concept of conditional
probabilities, it is possible to deduce the statistical operator of the system
after a measurement with a given result, which gives the probability
distribution for all possible consecutive measurements on the system. This
statistical operator, representing the state of the system after the first
measurement, is in general not the same that would be obtained using the
postulate of collapse.

\end{abstract}
\date{November 2007}
\maketitle

\section{Introduction}

As the measuring instruments are formed by the same kind of matter than
everything else, it seems natural to describe the measurement process by
quantum theory \cite{1}, \cite{1bis}. This was not the approach of Bohr, who
understood the measurement as a primitive notion, having a purely classical
description \cite{2}. The first attempt to use quantum theory to investigate
the measurement process was due to von Neumann \cite{3}. The quantum
interaction establishes a correlation between the macroscopic pointer
variables of the apparatus and the microscopic variables of the measured
system. In general the final state of the composed system obtained using the
Schr\"{o}dinger equation is a linear superposition of macroscopically
distinguishable values of the pointer variable. For those who interpret that
this state represents an instrument having simultaneously different pointer
positions, it is not clear how to relate this final composed state with the
definite pointer position that is perceived as a result of an actual single
measurement. This difficulty is generally named \textquotedblright the
measurement problem\textquotedblright. The collapse of the state vector,
either postulated or obtained from the addition of non linear terms to the
Schr\"{o}dinger equation, was an attempt to solve this problem. L. Ballentine
\cite{6} pointed out the inconsistencies of the collapse postulate with the
predictions of ordinary quantum theory, and a recent paper by M. Schlosshawer
\cite{7} discuss how ordinary quantum mechanics, with decoherence, can be
successfully used to avoid the addition of non linear terms to Schr\"{o}dinger
equation. Moreover, N. G. Van Kampen \cite{VK} and latter G. Sewell \cite{SEW}
stressed the importance of the macroscopic character of the measurement
instrument to deal with the measurement problem.

We do not try in this paper to give a solution to the ''measurement problem''
modifying the Schr\"{o}dinger equation to produce some kind of collapse. On
the contrary, we intend to \textit{deduce} the state in which the system is
prepared after a measurement with a given result, from the usual quantum
formalism applied to the interaction system-apparatus.

A defined choice of the interpretation for the state vector is unavoidable to
make contact between the mathematics of quantum theory and the results of the
experiments. In this paper the states of the systems are considered as
probability distributions, and the state vector is the mathematical tool to
compute these probabilities with the Born rule \cite{Pe}, \cite{Ba},
\cite{deM}, \cite{Peres}. The probabilities, and therefore the state vectors,
are properties of an \textit{ensemble} of systems. By the law of large numbers
these probabilities are related to the frequencies of results for a big
\textit{assembly} of identically prepared experiments \cite{AJP}. Moreover, in
this interpretation, the defined values of individual measurements are assumed
as primitive notions.

In section II we deduce de collapse of the wave function for the case of ideal
measurements. In section III we consider non ideal measurements and we show
that the collapse postulate is not verified. In section IV we deduce the
defining properties of a generalized measurement from considering the
measurement as a quantum process. The macroscopic character of the measurement
instrument was considered in section V. In appendix A we give a short
description of the logic of the measurement instruments, which is used through
the paper to describe probabilities for consecutive measurements.

\section{Ideal measurements and collapse}

The ideal measurement of an observable $Q$ is an interaction between the
system $S$ and the instrument $A$, which is represented by the following
unitary transformation in the Hilbert space $\mathcal{H}_{S}\otimes
\mathcal{H}_{A}$%
\[
|\phi\rangle|a_{0}\rangle\longrightarrow\sum_{q}\langle q|\phi\rangle
|q\rangle|a_{q}\rangle,
\]
where $|q\rangle$ is an eigenvector of the operator $\widehat{Q}$ with
eigenvalue $q$, $|a_{0}\rangle$ is the initial state of the instrument $A$ and
$|a_{q}\rangle$ is the state of the instrument correlated with the state
$|q\rangle$ of the system. The states of the instrument are eigenvectors of a
pointer observable $\widehat{A}$ ($\widehat{A}|a_{q}\rangle=a_{q}|a_{q}%
\rangle$, $\widehat{A}:\mathcal{H}_{A}\rightarrow\mathcal{H}_{A}$). For
simplicity we have not explicitly included in the description the huge number
of microscopic variables which together with the pointer define the state of
the measurement instrument. This case will be considered in section 5.

The ideal measurement of another observable $R$ requires a different
instrument $B$, and it is represented by a transformation in the corresponding
space $\mathcal{H}_{S}\otimes\mathcal{H}_{B}$%
\[
|\phi\rangle|b_{0}\rangle\longrightarrow\sum_{r}\langle r|\phi\rangle
|r\rangle|b_{r}\rangle,
\]
where $|r\rangle$ is an eigenvector of the operator $\widehat{R}$ with
eigenvalue $r$, $|b_{0}\rangle$ is the initial state of the instrument $B$ and
$|b_{r}\rangle$ is the state of the instrument correlated with the state
$|r\rangle$ of the system. The states of the instrument are eigenvectors of a
pointer observable $\widehat{B}$ ($\widehat{B}|b_{r}\rangle=b_{r}|b_{r}%
\rangle$, $\widehat{B}:\mathcal{H}_{B}\rightarrow\mathcal{H}_{B}$)

The consecutive measurements of the observables $Q$ and $R$ are represented by
consecutive transformations in the composed Hilbert space $\mathcal{H}$ of the
system $S$ and instruments $A$ and $B$ ($\mathcal{H}=\mathcal{H}_{S}%
\otimes\mathcal{H}_{A}\otimes\mathcal{H}_{B}$),
\begin{align}
|\Psi_{initial}\rangle &  =|\phi\rangle|a_{0}\rangle|b_{0}\rangle\nonumber\\
&  \longrightarrow\sum_{q}\langle q|\phi\rangle\,|q\rangle\,|a_{q}%
\rangle|b_{0}\rangle=\sum_{r}\sum_{q}\langle q|\phi\rangle\langle
r\,|q\rangle|r\rangle\,|a_{q}\rangle|b_{0}\rangle\label{2}\\
&  \longrightarrow\sum_{r}\sum_{q}\langle q|\phi\rangle\langle r\,|q\rangle
|r\rangle\,|a_{q}\rangle|b_{r}\rangle=|\Psi_{final}\rangle\nonumber
\end{align}

The propositions of a classical logic have the structure of an
orthocomplemented and distributive lattice \cite{Coh}. A classical logic can
be obtained for the propositions involving the pointer positions of both
measurement instruments. For these propositions the usual expressions of the
theory of probabilities are valid, particularly those corresponding to
conditional probabilities (see appendix). The use of conditional probabilities
to obtain the state of a system prepared by a measurement was given by W. M.
de Muynck (see section 3.3.4 of reference \cite{deM}).

The probability of measuring the value $r$ of the observable $R$ with the
second instrument $B$ , \textit{conditional }on having obtained the value $q$
of the observable $Q$ with the first instrument $A$, is given by
\begin{equation}
\Pr(b_{r}|a_{q})=\frac{\Pr(b_{r}\wedge a_{q})}{\Pr(a_{q})}=\frac{\langle
\Psi_{final}|(\widehat{I}_{S}\otimes|a_{q}\rangle\langle a_{q}|\otimes
|b_{r}\rangle\langle b_{r}|)|\Psi_{final}\rangle}{\langle\Psi_{final}%
|(\widehat{I}_{S}\otimes|a_{q}\rangle\langle a_{q}|\otimes\widehat{I}%
_{B})|\Psi_{final}\rangle}, \label{3}%
\end{equation}
where $\widehat{I}_{S}$ and $\widehat{I}_{B}$ are the identity operators in
the Hilbert spaces $\mathcal{H}_{S}$ and $\mathcal{H}_{B}$ and we have used
the Born rule for computing the probabilities $\Pr(b_{r}\wedge a_{q})$ and
$\Pr(a_{q})$. Taking into account the expression for the final state given by
equation (\ref{2}), it is straightforward to prove that the conditional
probability given by equation (\ref{3}) can be written in the following simple
way
\[
\Pr(b_{r}|a_{q})=\langle q|r\rangle\langle r|q\rangle.
\]

Moreover, if we consider the projector operator $\widehat{\Pi}_{r}%
\equiv|r\rangle\langle r|$ corresponding to the proposition $r=R$, and if we
\textit{define} $\widehat{\rho}_{q}\equiv|q\rangle\langle q|$, the conditional
probability can be given the expression
\begin{equation}
\Pr(b_{r}|a_{q})=Tr[\widehat{\rho}_{q}\widehat{\Pi}_{r}]. \label{4}%
\end{equation}

The first term refer to the probability of certain values of the pointer
positions of the instruments $A$ and $B$, while the second term is written in
terms of vectors and operators of the Hilbert space $\mathcal{H}_{S}$ of the
system $S$.

If we perform a different sequence of measurements on the system, maintaining
the first instrument measuring the observable $Q$, but changing the second
instrument for one suitable to the ideal measurement of the observable
$R^{\prime}$, we will obtain
\begin{equation}
\Pr(b_{r}^{\prime}|a_{q})=Tr[\widehat{\rho}_{q}\widehat{\Pi}_{r}^{\prime}],
\label{5}%
\end{equation}
where $\widehat{\Pi}_{r}^{\prime}\equiv|r^{\prime}\rangle\langle r^{\prime}|$
is the projector corresponding to the proposition $R^{\prime}=r^{\prime}$.

Eqs. (\ref{4}) and (\ref{5}) give the probabilities to obtain the result $r$
for the measurement of the observable $R$ and the result $r^{\prime}$ for the
observable $R^{\prime}$, respectively. Therefore, the presence of the
corresponding projectors $\widehat{\Pi}_{r}$ and $\widehat{\Pi}_{r}^{\prime}$
in the second terms. Moreover, in both cases, the probabilities are
\textit{conditional} to have previously obtained the result $q $ from the
measurement of the observable $Q$. In other words, in both cases the
measurements of $R$ and $R^{\prime}$ are performed on an ensemble of systems
$S$ for which the result $q$ of the observable $Q$ was previously obtained.

Eqs. (\ref{4}) and (\ref{5}) also show that this special ensemble of systems
is represented by the state operator $\widehat{\rho}_{q}\equiv|q\rangle\langle
q|$. It is evident that this state operator is suitable to compute the
probabilities for the values of \textit{any} observable of the system, for the
ensemble of systems in which a previous ideal measurement of the observable
$Q$ has given the value $q$.

The initial state of the system is represented by the vector $|\phi\rangle
\in\mathcal{H}_{S}$, while after the measurement it is represented by the
vector $|q\rangle\in\mathcal{H}_{S}$, the eigenvector of the operator
$\widehat{Q}$ with eigenvalue $q$. This result would also have been obtained
by using the collapse postulate.

However we did not use the collapse postulate to obtain the result. It was
obtained using i) Schr\"{o}dinger equation for the unitary evolution given in
eq. (\ref{2}) of the state vector corresponding to the closed system formed by
the system $S$ and the instruments $A$ and $B$, and ii) conditional
probability \textit{defined} by eq. (\ref{3}) as a quotient of probabilities
obtained from the Born rule.

The transformation $|\phi\rangle\rightarrow|q\rangle$ of the state of the
system $S$ due to the measurement has some remarkable properties which make it
very different from the transformations generated by the Schr\"{o}dinger equation:

i) it is not a unitary transformation (different states $|\phi\rangle$ and
$|\phi^{\prime}\rangle$ may evolve into the same state $|q\rangle$)

ii) the transformation $|\phi\rangle\rightarrow|q\rangle$ do not represent the
evolution of a single ensemble of systems ($|q\rangle$ represents the state of
a subensemble of the ensemble unitarily evolved from the state $|\phi\rangle$).

For the case of an ideal measurement, this transformation coincides with the
one provided by the collapse postulate, but we have avoided to use this
postulate. In our approach the measurement is analyzed as a process fully
described by quantum theory. The non-unitary transformation $|\phi
\rangle\rightarrow|q\rangle$ was deduced from the unitary evolution generated
by the Schr\"{o}dinger equation describing the interaction system-apparatus.

The case of an ideal measurement of an observable with degenerate spectrum can
also be obtained in this approach. Let us consider an observable represented
by the operator%

\[
\widehat{Q}=\sum_{q}q\,\widehat{\Pi}_{q},\qquad\widehat{\Pi}_{q}=\sum
_{j=1}^{n_{q}}|q,j\rangle\langle q,j|,\qquad\langle q,j|q^{\prime},j^{\prime
}\rangle=\delta_{qq^{\prime}}\delta_{jj^{\prime}},
\]
where $n_{q}$ is the dimension of the subspace of $\mathcal{H}_{S}$
corresponding to the eigenvectors of $\widehat{Q}$ with eigenvalue $q$. Any
vector $|\phi\rangle\in\mathcal{H}_{S}$ can be written in terms of the
projectors $\widehat{\Pi}_{q}$%
\[
|\phi\rangle=\sum_{q}\sum_{j=1}^{n_{q}}c_{qj}|q,j\rangle=\sum_{q}\widehat{\Pi
}_{q}|\phi\rangle,
\]
where $c_{qj}\equiv\langle q,j|\phi\rangle$.

An ideal measurement of this observable by an instrument $A$ is represented by
the following unitary transformation in $\mathcal{H}_{S}\otimes\mathcal{H}%
_{A}$%
\[
|q,j\rangle|a_{0}\rangle\longrightarrow|q,j\rangle|a_{q}\rangle.
\]

After the interaction with instrument $A$, the system $S$ interacts with
another instrument $B$, making an ideal measurement of an observable
represented by the operator $\widehat{R}=\sum_{r}r|r\rangle\langle r|$, having
non degenerate spectrum. The second measurement is represented by the
transformation $|r\rangle|b_{0}\rangle\rightarrow|r\rangle|b_{r}\rangle$.

The consecutive measurements are represented by an unitary transformation in
$\mathcal{H=H}_{S}\otimes\mathcal{H}_{A}\otimes\mathcal{H}_{B}$%
\begin{align*}
|\Psi_{initial}\rangle &  =|\phi\rangle|a_{0}\rangle|b_{0}\rangle
\longrightarrow\sum_{q}\widehat{\Pi}_{q}|\phi\rangle|a_{q}\rangle|b_{0}%
\rangle\\
&  \longrightarrow\sum_{q}\sum_{r}|r\rangle\langle r|\widehat{\Pi}_{q}%
|\phi\rangle|a_{q}\rangle|b_{r}\rangle.
\end{align*}

For the probability to obtain $b_{r}$ in the second measurement if the the
result of the first one was $a_{q}$ we obtain in this case%
\[
\Pr(b_{r}|a_{q})=\frac{\Pr(b_{r}\wedge a_{q})}{\Pr(a_{q})}=Tr[\widehat{\rho
}_{q}\widehat{\Pi}_{r}],
\]
where $\widehat{\Pi}_{r}=|r\rangle\langle r|$ and
\[
\widehat{\rho}_{q}=\frac{\widehat{\Pi}_{q}|\phi\rangle\langle\phi|\widehat
{\Pi}_{q}}{\langle\phi|\widehat{\Pi}_{q}|\phi\rangle},
\]
which is the L\"{u}ders projection.

\section{Non ideal measurements}

In this case the system is modified by the measurement process, even when the
initial state of the system is an eigenstate of the observable to be measured.

The measurement processes on the eigenvectors $|q\rangle$ and $|r\rangle$ of
the operators $\widehat{Q}$ and $\widehat{R}$ are described by the following
unitary transformations
\[
|q\rangle|a_{0}\rangle\rightarrow|\mu_{q}\rangle|a_{q}\rangle,\qquad
|r\rangle|b_{0}\rangle\rightarrow|\nu_{r}\rangle|b_{r}\rangle,
\]
where $|\mu_{q}\rangle$ and $|\nu_{r}\rangle$ are different from the initial
states $|q\rangle$ and $|r\rangle$.

Consecutive measurements are represented by the transformation
\begin{align*}
|\Psi_{initial}\rangle &  =|\phi\rangle|a_{0}\rangle|b_{0}\rangle
\longrightarrow\sum_{q}\langle q|\phi\rangle\,|\mu_{q}\rangle\,|a_{q}%
\rangle|b_{0}\rangle=\\
&  =\sum_{r}\sum_{q}\langle q|\phi\rangle\langle r\,|\mu_{q}\rangle
|r\rangle\,|a_{q}\rangle|b_{0}\rangle\\
&  \longrightarrow\sum_{r}\sum_{q}\langle q|\phi\rangle\langle r\,|\mu
_{q}\rangle|\nu_{r}\rangle\,|a_{q}\rangle|b_{r}\rangle=|\Psi_{final}\rangle,
\end{align*}
and the probability that the second instrument measures the value $r$ of $R$
\textit{if} the first instrument has measured the value $q$ of $Q$ is given by
the conditional probability
\begin{align*}
\Pr(b_{r}|a_{q})  &  =\frac{\Pr(b_{r}\wedge a_{q})}{\Pr(a_{q})}=\frac
{\langle\Psi_{final}|(\widehat{I}_{S}\otimes|a_{q}\rangle\langle a_{q}%
|\otimes|b_{r}\rangle\langle b_{r}|)|\Psi_{final}\rangle}{\langle\Psi
_{final}|(\widehat{I}_{S}\otimes|a_{q}\rangle\langle a_{q}|\otimes\widehat
{I}_{B})|\Psi_{final}\rangle}=\\
&  =\frac{|\langle q|\phi\rangle|^{2}|\langle r\,|\mu_{q}\rangle|^{2}}%
{\sum_{r}|\langle q|\phi\rangle|^{2}|\langle r\,|\mu_{q}\rangle|^{2}}=|\langle
r\,|\mu_{q}\rangle|^{2}=Tr[\widehat{\rho}_{q}^{\prime}\widehat{\Pi}_{r}],
\end{align*}
where $\widehat{\Pi}_{r}=|r\rangle\langle r|$ corresponds to the proposition
$r=R$, and we \textit{define} $\widehat{\rho}_{q}^{\prime}\equiv|\mu
_{q}\rangle\langle\mu_{q}|$.

In this case we have shown that the first measurement with result $q$ has
prepared the system in the state $\widehat{\rho}_{q}^{\prime}$. The effect of
the first measurement on the system is in this case the transformation
$|\phi\rangle\rightarrow|\mu_{q}\rangle$, which do not coincide with the
collapse postulate. This result was previously obtained by L. E. Ballentine
\cite{6}, who analyzed the limitations of the collapse postulate.

\section{Generalized measurements}

Now we consider the most general measurement process \cite{Chuang}. It is
described through a collection of \textit{measurement operators}
$\{\widehat{M}_{m}\}$, acting on the Hilbert space $\mathcal{H}_{S}$ of the
system, and satisfying $\sum_{m}\widehat{M}_{m}^{\dagger}\widehat{M}%
_{m}=\widehat{I}_{S}$. The probability to obtain the result $m$ in the
measurement on a state $|\phi\rangle$ is $\Pr(m)=\langle\phi|\widehat{M}%
_{m}^{\dagger}\widehat{M}_{m}|\phi\rangle$, and if the result is $m$ the
transformation on the system is
\begin{equation}
|\phi\rangle\rightarrow\left(  \sqrt{\langle\phi|\widehat{M}_{m}^{\dagger
}\widehat{M}_{m}|\phi\rangle}\right)  ^{-1}\widehat{M}_{m}|\phi\rangle.
\label{6}%
\end{equation}

In this section we are going to prove that these \textit{defining properties}
of a generalized measurement can be \textit{deduced} considering the
interaction between the system $S$ and a measurement instrument $A$,
represented by a unitary transformation $\widehat{U}$ in the Hilbert space
$\mathcal{H}=\mathcal{H}_{S}\otimes\mathcal{H}_{A}$. If we denote by
$|m\rangle$ the state of the instrument corresponding to the result $m$, the
measurement operators can be deduced from the following expression
\begin{align*}
|\Psi_{initial}\rangle &  =|\phi\rangle|0\rangle\rightarrow|\Psi
_{final}\rangle=\widehat{U}(|\phi\rangle|0\rangle)=\\
&  =\sum_{m}|m\rangle\langle m|\widehat{U}(|\phi\rangle|0\rangle)\equiv
\sum_{m}(\widehat{M}_{m}|\phi\rangle)|m\rangle.
\end{align*}

The probability to obtain the result $m$ can be deduced from the Born rule
\[
\Pr(m)=\langle\Psi_{final}|(\widehat{I}_{S}\otimes|m\rangle\langle
m|)|\Psi_{final}\rangle=\langle\phi|\widehat{M}_{m}^{\dagger}\widehat{M}%
_{m}|\phi\rangle.
\]

If two instruments $A$ and $B$, with measurement operators $\{\widehat
{M}_{m_{A}}\}$ and $\{\widehat{N}_{m_{B}}\}$, are used for consecutive
measurements on a system $S$, the process is represented by the following two
consecutive unitary transformations
\begin{align*}
|\Psi_{initial}\rangle &  =|\phi\rangle|0_{A}\rangle|0_{B}\rangle
\rightarrow\sum_{m_{A}}(\widehat{M}_{m_{A}}|\phi\rangle)|m_{A}\rangle
|0_{B}\rangle\\
&  \rightarrow\sum_{m_{A}}\sum_{m_{B}}(\widehat{N}_{m_{B}}\widehat{M}_{m_{A}%
}|\phi\rangle)|m_{A}\rangle|m_{B}\rangle=|\Psi_{final}\rangle.
\end{align*}

The probability for the instrument $B$ to give the result $m_{B}$ conditioned
for the fact that the instrument $A$ has already given the result $m_{A}$ is
now obtained from the expression of conditional probability and the Born rule
\begin{align*}
\Pr(m_{B}|m_{A})  &  =\frac{\Pr(m_{B}\wedge m_{A})}{\Pr(m_{A})}=\\
&  =\frac{\langle\Psi_{final}|(\widehat{I}_{S}\otimes|m_{A}\rangle\langle
m_{A}|\otimes|m_{B}\rangle\langle m_{B}|)|\Psi_{final}\rangle}{\langle
\Psi_{final}|(\widehat{I}_{S}\otimes|m_{A}\rangle\langle m_{A}|\otimes
\widehat{I}_{B})|\Psi_{final}\rangle}=\\
&  =\langle\phi_{m_{A}}|\widehat{N}_{m_{B}}^{\dagger}\widehat{N}_{m_{B}}%
|\phi_{m_{A}}\rangle,\\
|\phi_{m_{A}}\rangle &  \equiv\left(  \sqrt{\langle\phi|\widehat{M}_{m_{A}%
}^{\dagger}\widehat{M}_{m_{A}}|\phi\rangle}\right)  ^{-1}\widehat{M}_{m_{A}%
}|\phi\rangle
\end{align*}

The state $|\phi_{m_{A}}\rangle$ can be interpreted as the result of a
preparation on the system produced when the instrument $A$ registers the value
$m_{A}$. The postulated generalized collapse defined in eq.(\ref{6}) is now
deduced from Schr\"{o}dinger equation and Born rule, by considering the
measurement instruments as quantum systems. In this section we have shown,
once again, that all the properties defining a general measurement can be
deduced considering the measurement as a quantum process of interaction
between system and instruments, and that there is no need of collapse postulate.

\section{Macroscopic instruments}

In the previous sections we have not included the huge number of microscopic
variables of the macroscopic measurement instrument. Including these
variables, an operator $\widehat{A}$ representing the pointer of an instrument
$M_{1}$ has a complete set of eigenvectors in the Hilbert space $\mathcal{H}%
_{M_{1}}$, satisfying $\widehat{A}|a,m\rangle=a|a,m\rangle$, where $a$ is the
pointer variable, and $m$ labels the many other quantum numbers necessary to
specify an eigenvector. For the system $S$, we consider the measurement of an
observable represented by an operator $\widehat{Q}$ in the Hilbert space
$\mathcal{H}_{S}$, having a complete set of eigenvectors verifying
$\widehat{Q}|q\rangle=q|q\rangle$.

The non ideal measurement process is represented by an unitary transformation
in the Hilbert space $\mathcal{H}_{S}\otimes\mathcal{H}_{M_{1}}$, defined by%
\[
|q\rangle|a_{0},m\rangle\longrightarrow|a_{q};(q,m)\rangle\equiv%
%TCIMACRO{\dsum \limits_{q^{\prime}m^{\prime}}}%
%BeginExpansion
{\displaystyle\sum\limits_{q^{\prime}m^{\prime}}}
%EndExpansion
u_{qm}^{q^{\prime}m^{\prime}}|q^{\prime}\rangle|a_{q},m^{\prime}\rangle.
\]

Following L. E. Ballentine \cite{6}, the labels $(r,m)$ in the final vector do
not denote eigenvalues, but they keep the memory of the initial state previous
to the measurement. The system-instrument state after the measurement is
$|a_{q};(q,m)\rangle$, having a well defined value $a_{q}$ of the pointer
variable, but in general not well defined values of the remaining variables.
The \textit{initial} value $q$ of the system observable $Q$ is correlated with
the \textit{final} value $a_{q}$ of the pointer observable $A$.

Analogously, the measurement of another observable represented by the operator
$\widehat{R}$ in the Hilbert space $\mathcal{H}_{S}$ of the system $S$, is
made with an instrument $M_{2}$ with pointer operator $\widehat{B}$ in the
Hilbert space $\mathcal{H}_{M_{2}}$. The measurement process is represented by
an unitary transformation in the Hilbert space $\mathcal{H}_{S}\otimes
\mathcal{H}_{M_{2}}$%
\[
|r\rangle|b_{0},n\rangle\longrightarrow|b_{r};(r,n)\rangle\equiv%
%TCIMACRO{\dsum \limits_{r^{\prime}n^{\prime}}}%
%BeginExpansion
{\displaystyle\sum\limits_{r^{\prime}n^{\prime}}}
%EndExpansion
v_{rn}^{r^{\prime}n^{\prime}}|r^{\prime}\rangle|b_{r},n^{\prime}\rangle,
\]
where $|r\rangle$ is an eigenvector of $\widehat{R}$ in $\mathcal{H}_{S}$
($\widehat{R}|r\rangle=r|r\rangle$) and $|b,n\rangle$ is an eigenvector of the
pointer observable $\widehat{B}$ in the Hilbert space $\mathcal{H}_{B}$
($\widehat{B}|b,n\rangle=b|b,n\rangle$). The index $n$ represents the quantum
numbers different from the label $b$ associated to the pointer.

For an initial state $|\phi\rangle=\sum\nolimits_{q}c_{q}|q\rangle$ of the
system $S$ ($c_{q}\equiv\langle q|\phi\rangle$), the consecutive measurement
of observables $\widehat{Q}$ and $\widehat{R}$ is represented by the following
consecutive transformation in the Hilbert space $\mathcal{H}_{S}%
\otimes\mathcal{H}_{M_{1}}\otimes\mathcal{H}_{M_{2}}$%
\begin{align*}
|\Psi_{initial}\rangle &  =|\phi\rangle|a_{0};m\rangle|b_{0};n\rangle\\
&  \longrightarrow\sum_{q}c_{q}|a_{q};(q,m)\rangle|b_{0},n\rangle\\
&  \longrightarrow|\Psi_{final}\rangle\equiv\sum_{q}c_{q}\sum_{r}%
(r|a_{q};(q,m))|b_{r};(r,n)\rangle,
\end{align*}
where $(r|a_{q};(q,m))\equiv\sum_{q^{\prime}m^{\prime}}u_{qm}^{q^{\prime
}m^{\prime}}\langle r|q^{\prime}\rangle|a_{q},m^{\prime}\rangle\in
\mathcal{H}_{M_{1}}$.

By straightfordward calculations we obtain%
\begin{align*}
\Pr(b_{r}|a_{q})  &  =\frac{\Pr(b_{r}\wedge a_{q})}{\Pr(a_{q})}\\
&  =\frac{\langle\Psi_{final}|\left[  \widehat{I}_{S}\otimes\sum_{m^{\prime}%
}|a_{q},m^{\prime}\rangle\langle a_{q},m^{\prime}|\otimes\sum_{n^{\prime}%
}|b_{r},n^{\prime}\rangle\langle b_{r},n^{\prime}|\right]  |\Psi
_{final}\rangle}{\langle\Psi_{final}|\left[  \widehat{I}_{S}\otimes
\sum_{m^{\prime}}|a_{q},m^{\prime}\rangle\langle a_{q},m^{\prime}%
|\otimes\widehat{I}_{M_{2}}\right]  |\Psi_{final}\rangle}\\
&  =Tr[\widehat{\rho}(q,m)\widehat{\Pi}_{r}],
\end{align*}
where $\widehat{\Pi}_{r}\equiv|r\rangle\langle r|$ and
\[
\widehat{\rho}(q,m)\equiv\sum_{q^{\prime}q^{\prime\prime}}\left(
\sum_{\widetilde{m}}u_{qm}^{q^{\prime}\widetilde{m}}\overline{u}%
_{qm}^{q^{\prime\prime}\widetilde{m}}\right)  |q^{\prime}\rangle\langle
q^{\prime\prime}|.
\]

The density operator $\widehat{\rho}(q,m)$ represents the state of the system
$S$ after the measurement with the instrument $M_{1}$ has given the result
$a_{q}$. We notice in this case an important difference with the results
obtained in the previous sections: even for a system $S$ in an initially pure
state, the effect of the instrument microscopic variables is to prepare the
system in a non pure state.

\section{Conclusions}

The collapse of the wave function is usually invoqued to justify the existence
of a well defined result of a single measurement process.

Our strategy in this paper has been the opposite. First, we gave a full
quantum description of the system- instrument interaction for the measurement
process. Second, we accepted the experimental evidence that in each individual
experiment, the measurement instrument produce a well defined results. Third,
we obtained the probabilities for these results using the Born rule.

For two consecutive measurements on the system, the probability distribution
of the possible results of the second measurement conditioned to a determined
result of the first one, can be computed with the usual expression for
conditional probabilities. From this calculations we have been able to deduce
which is the state vector representing the system after a measurement with a
given result.

The system is \textit{prepared} in a well defined state by the measurement.
This state is strongly dependent on the form of the interaction
system-apparatus. The obtained result coincides with that of the collapse
postulate only for the ideal measurement, and explicit expressions of the
prepared state for non ideal and generalized measurements have also been obtained.

In this way we have been able to provide a satisfactory description of the
measurement process as a quantum process, in which it is not necessary to
postulate additional physical mechanisms like the collapse of the wave function.

\appendix

\section{The logic of the measurement instruments}

Several times in this paper we have considered the measurement of an
observables $Q$ with an instrument $A$ on a system $S$, followed by the
measurement of another observable $R$ using a second instrument $B$. The whole
process was described by the evolution of a state vector in the Hilbert space
$\mathcal{H}=\mathcal{H}_{S}\otimes\mathcal{H}_{A}\otimes\mathcal{H}_{B}$. We
labelled by $\widehat{A}$ and $\widehat{B}$ the corresponding pointer
operators having eigenvalues $a_{q}$ and $b_{p}$, and eigenvectors
$|a_{q}\rangle$ and $|b_{p}\rangle$.

The quantum description of the measurement process should prescribe definite
values for the probabilities of propositions like ''the result on the first
instrument was $a_{q}$ \textit{and} the result on the second instrument is
$b_{p}$'', or ''the result on the second instrument is $b_{p}$ \textit{if} the
result on the first instrument was $a_{q}$''. These propositions involve
eigenvalues of the pointer operators $\widehat{A}$ and $\widehat{B}$, acting
on Hilbert spaces $\mathcal{H}_{A}$ and $\mathcal{H}_{B}$. These operators can
be lifted to operators acting on the tensor product space $\mathcal{H},$%
\[
\widehat{A}_{\mathcal{H}}\equiv\widehat{I}_{S}\otimes\widehat{A}%
\otimes\widehat{I}_{B},\qquad\widehat{B}_{\mathcal{H}}\equiv\widehat{I}%
_{S}\otimes\widehat{I}_{A}\otimes\widehat{B},
\]
where $\widehat{I}_{S}$, $\widehat{I}_{A}$ and $\widehat{I}_{B}$ are the
identity operators in the spaces $H_{S}$, $H_{A}$ and $H_{B}$. It is evident
that the lifted operators $\widehat{A}_{\mathcal{H}}$ and $\widehat
{B}_{\mathcal{H}}$ commute, and therefore the possible results of the
consecutive measurements have the quantum logic of the simultaneous
eigenvectors of a set of commuting operators. The relevant aspect of this
logic are reviewed in what follows.

Let us consider a complete set of commuting observables, represented by $n$
operators $\widehat{\overline{R}}\equiv(\widehat{R}_{1},...,\widehat{R}_{n})
$, having the complete orthonormal eigenvectors $|\overline{r}\rangle
=|r_{1},...,r_{n}\rangle$ ($\widehat{\overline{R}}\,|\overline{r}%
\rangle=\overline{r}\,|\overline{r}\rangle$, $\overline{r}\in\mathbb{R}^{n}$).
The proposition ''$\overline{r}$ belongs to the set $\Delta^{n}\subset
\mathbb{R}^{n}$'' is represented by the subspace of the Hilbert space
generated by the projector $\widehat{\Pi}_{\Delta^{n}}=\sum_{\overline{r}%
\in\Delta^{n}}|\overline{r}\rangle\langle\overline{r}|$. The conjunction and
disjunction of two proposition are represented by the intersection and the
direct sum of the subspaces. The order relation is the implication,
represented by set inclusion. For two propositions $p_{1}=\{\overline{r}%
\in\Delta_{1}^{n}\}$ and $p_{2}=\{\overline{r}\in\Delta_{2}^{n}\}$ we have the
following corresponding projectors \cite{Mit},
\begin{equation}%
\begin{array}
[c]{lll}%
p_{1} & \longrightarrow & \widehat{\Pi}_{1}=\sum_{\overline{r}\in\Delta
_{1}^{n}}|\overline{r}\rangle\langle\overline{r}|\\
p_{2} & \longrightarrow & \widehat{\Pi}_{2}=\sum_{\overline{r}\in\Delta
_{2}^{n}}|\overline{r}\rangle\langle\overline{r}|\\
p_{1}\wedge p_{2} & \longrightarrow & \lim_{k\rightarrow\infty}(\widehat{\Pi
}_{1}\widehat{\Pi}_{2})^{k}\\
p_{1}\vee p_{2} & \longrightarrow & \widehat{I}-\lim_{k\rightarrow\infty
}[(\widehat{I}-\widehat{\Pi}_{1})(\widehat{I}-\widehat{\Pi}_{2})]^{k}\\
p_{1}^{\prime} & \longrightarrow & \widehat{I}-\widehat{\Pi}_{1}%
\end{array}
\label{A1}%
\end{equation}

The projectors associated with propositions within the basis $\{|\overline
{r}\rangle\}$ are commutative
\[
\widehat{\Pi}_{1}\widehat{\Pi}_{2}=\sum_{\overline{r}\in\Delta_{1}^{n}%
}|\overline{r}\rangle\langle\overline{r}|\sum_{\overline{r}^{\prime}\in
\Delta_{2}}|\overline{r}^{\prime}\rangle\langle\overline{r}^{\prime}%
|=\sum_{\overline{r}\in\Delta_{1}^{n}\frown\Delta_{2}^{n}}|\overline{r}%
\rangle\langle\overline{r}|=\widehat{\Pi}_{2}\widehat{\Pi}_{1}%
\]

From these commutation properties simplified expressions are easily obtained
for the projectors associated with conjunction and disjunction
\[%
\begin{array}
[c]{lll}%
p_{1}\wedge p_{2} & \longrightarrow & \widehat{\Pi}_{1}\widehat{\Pi}_{2}\\
p_{1}\vee p_{2} & \longrightarrow & \widehat{\Pi}_{1}+\widehat{\Pi}%
_{2}-\widehat{\Pi}_{1}\widehat{\Pi}_{2}%
\end{array}
\]

Propositions of the form $p_{1}=\{\overline{r}\in\Delta_{1}^{n}\}$,
$p_{2}=\{\overline{r}\in\Delta_{2}^{n}\}$ and $p_{3}=\{\overline{r}\in
\Delta_{3}^{n}\}$ are distributive, i.e.
\begin{align*}
p_{1}\wedge(p_{2}\vee p_{3})  &  =(p_{1}\wedge p_{2})\vee(p_{1}\wedge
p_{3}),\\
p_{1}\vee(p_{2}\wedge p_{3})  &  =(p_{1}\vee p_{2})\wedge(p_{1}\vee p_{3}),
\end{align*}
as can be easily proved by writing the corresponding projectors. Therefore,
within a fixed basis, the lattice of propositions is a classical logic.
Moreover, within a fixed basis the usual logic of our language is suitable to
talk about quantum propositions.

A probability distribution on a lattice is a function from the propositions to
the real numbers satisfying

i) $\Pr(p)\geqslant0$, for all propositions $p$

ii) $\Pr(p\vee q)=\Pr(p)+\Pr(q)$ for all propositions $p$ and $q$ such that
$p\wedge q=\phi$

iii) $\Pr(I)=1$ for the unit proposition $I.$

Probabilities in quantum theory are calculated using the Born rule. For a pure
state represented by the vector $\psi$ of the Hilbert space, the probability
of a proposition $p$ is given by $\Pr(p)=\langle\psi|\widehat{\Pi}_{p}%
|\psi\rangle$, where $\widehat{\Pi}_{p}$ is the projector associated with the
proposition $p$. We can prove that conditions i) ii) and iii) are satisfied.

To prove condition i) consider a proposition $p_{\Delta^{n}}=\{\overline{r}%
\in\Delta^{n}\}$, with the corresponding projector $\widehat{\Pi}_{\Delta^{n}%
}=\sum_{\overline{r}\in\Delta^{n}}|\overline{r}\rangle\langle\overline{r}|$
and compute $\Pr(p_{\Delta^{n}})=\langle\psi|\widehat{\Pi}_{\Delta^{n}}%
|\psi\rangle=\sum_{\overline{r}\in\Delta^{n}}\langle\psi|\overline{r}%
\rangle\langle\overline{r}|\psi\rangle=\sum_{\overline{r}\in\Delta^{n}%
}|\langle\overline{r}|\psi\rangle|^{2}\geqslant0$

To prove ii) let us consider two disjoint subsets $\Delta_{1}^{n}$ and
$\Delta_{2}^{n}$of $\mathbb{R}^{n}$. Therefore $\widehat{\Pi}_{\Delta_{1}^{n}%
}\widehat{\Pi}_{\Delta_{2}^{n}}=0$, and therefore $p_{\Delta_{1}^{n}}\wedge
p_{\Delta_{2}^{n}}=\phi$. The projector corresponding to the proposition
$p_{\Delta_{1}^{n}}\vee p_{\Delta_{2}^{n}}$ is $\widehat{\Pi}_{\Delta_{1}^{n}%
}+\widehat{\Pi}_{\Delta_{2}^{n}}-\widehat{\Pi}_{\Delta_{1}^{n}}\widehat{\Pi
}_{\Delta_{2}^{n}}=\widehat{\Pi}_{\Delta_{1}^{n}}+\widehat{\Pi}_{\Delta
_{2}^{n}}$. The probability of the disjunction is
\begin{align*}
\Pr(p_{\Delta_{1}^{n}}\vee p_{\Delta_{2}^{n}})  &  =\langle\psi|(\widehat{\Pi
}_{\Delta_{1}^{n}}+\widehat{\Pi}_{\Delta_{2}^{n}})|\psi\rangle=\\
&  =\langle\psi|\widehat{\Pi}_{\Delta_{1}^{n}}|\psi\rangle+\langle
\psi|\widehat{\Pi}_{\Delta_{2}^{n}}|\psi\rangle=\Pr(p_{\Delta_{1}^{n}}%
)+\Pr(p_{\Delta_{2}^{n}}),
\end{align*}
and condition ii) is verified.

Property iii) is easily obtained
\[
\Pr(p_{\mathbb{R}^{n}})=\langle\psi|\widehat{\Pi}_{\mathbb{R}^{n}}|\psi
\rangle=\sum_{\overline{r}\in\mathbb{R}^{n}}\langle\psi|\overline{r}%
\rangle\langle\overline{r}|\psi\rangle=\langle\psi|\widehat{I}|\psi\rangle=1
\]

The probability for the proposition \textquotedblright the observable $R_{j}$
has the value $r_{j}$ in the set $\Delta_{j}$ \textit{if }the observable
$R_{i}$ has the value $r_{i}$ in the set $\Delta_{i}$\textquotedblright\ can
be \textit{defined} by the standard expression for the conditional
probability
\begin{align}
\Pr(p_{\Delta_{j}}|p_{\Delta_{i}})  &  \equiv\frac{\Pr(p_{\Delta_{j}}\,\wedge
p_{\Delta_{i}})}{\Pr(p_{\Delta_{i}})},\nonumber\\
p_{\Delta_{j}}  &  \equiv\{r_{j}\in\Delta_{j}\subset\mathbb{R}\},\qquad
p_{\Delta_{i}}\equiv\{r_{i}\in\Delta_{i}\subset\mathbb{R}\}, \label{A3}%
\end{align}
which is well defined if $\Pr(p_{\Delta_{i}})\neq0$. To be consistent, we must
verify that the expression just defined satisfies the probability conditions
i) ii) and iii).

It is obvious that $\Pr(p_{\Delta_{j}}|p_{\Delta_{i}})\geqslant0$, and
therefore condition i) is verified.

Let us consider that $\Delta_{j}$ and $\Delta_{j}^{\prime}$ are two disjoint
subsets of $\mathbb{R}$ ($\Delta_{j}\cap\Delta_{j}^{\prime}=\phi$). Therefore
the propositions $p_{\Delta_{j}}\equiv\{r_{j}\in\Delta_{j}\}$ and
$p_{\Delta_{j}^{\prime}}\equiv\{r_{j}\in\Delta_{j}^{\prime}\}$satisfy
$p_{\Delta_{j}}\wedge p_{\Delta_{j}^{\prime}}=\phi$. Let us consider
\begin{align*}
\Pr(p_{\Delta_{j}}\vee p_{\Delta_{j}^{\prime}}|p_{\Delta_{i}})  &  \equiv
\frac{\Pr([p_{\Delta_{j}}\,\vee p_{\Delta_{j}^{\prime}}]\wedge p_{\Delta_{i}%
})}{\Pr(p_{\Delta_{i}})}=\\
&  =\frac{\Pr([p_{\Delta_{j}}\,\wedge p_{\Delta_{i}}]\vee[p_{\Delta
_{j}^{\prime}}\wedge p_{\Delta_{i}}])}{\Pr(p_{\Delta_{i}})}=\\
&  =\frac{\Pr(p_{\Delta_{j}}\,\wedge p_{\Delta_{i}})+\Pr(p_{\Delta_{j}%
^{\prime}}\wedge p_{\Delta_{i}})}{\Pr(p_{\Delta_{i}})},
\end{align*}
where the last term follows from the fact that
\[
\lbrack p_{\Delta_{j}}\,\wedge p_{\Delta_{i}}]\wedge[p_{\Delta_{j}^{\prime}%
}\wedge p_{\Delta_{i}}]=(p_{\Delta_{j}}\,\wedge p_{\Delta_{j}^{\prime}})\wedge
p_{\Delta_{i}}=\phi\wedge p_{\Delta_{i}}=\phi.
\]

Therefore $\Pr(p_{\Delta_{j}}\vee p_{\Delta_{j}^{\prime}}|p_{\Delta_{i}}%
)=\Pr(p_{\Delta_{j}}|p_{\Delta_{i}})+\Pr(p_{\Delta_{j}^{\prime}}|p_{\Delta
_{i}})$, and we have verified condition ii).

Condition iii) is easily verified, as it is self evident from the following
equation
\[
\Pr(p_{\mathbb{R}}|p_{\Delta_{i}})=\frac{\Pr(p_{\mathbb{R}}\wedge
p_{\Delta_{i}})}{\Pr(p_{\Delta_{i}})}=\frac{\Pr(p_{\mathbb{R\cap}\Delta_{i}}%
)}{\Pr(p_{\Delta_{i}})}=\frac{\Pr(p_{\Delta_{i}})}{\Pr(p_{\Delta_{i}})}=1.
\]

We emphasize that the consistency of the definition of the conditional
probability given in eq. (\ref{A3}) relies strongly on the fact that it is
applied to propositions within a fixed basis of the Hilbert space. This is
precisely the case in this paper, where we deal with propositions
corresponding to the possible results of consecutive measurements.

\end{document}